\documentclass[12pt]{article}

\usepackage{graphicx}
\usepackage{epsfig}
\usepackage{amsmath}
\usepackage{mathrsfs}
\usepackage{amssymb}
\usepackage{amsthm}
\usepackage[authoryear]{natbib}
\usepackage{algorithm,algorithmic}
\usepackage{url}
\usepackage[margin=1in,paperwidth=8.5in,paperheight=11in]{geometry}
\usepackage{rotating}
\usepackage{setspace}
\usepackage{stmaryrd}
\usepackage{stackrel}
\usepackage{hyperref}

\makeatletter
\newcommand\code{\bgroup\@makeother\_\@makeother\~\@makeother\$\@makeother\^\@codex}
\def\@codex#1{{\normalfont\ttfamily\hyphenchar\font=-1 #1}\egroup}

\newcommand\proglang{\bgroup\@makeother\_\@makeother\~\@makeother\$\@makeother\^\@codex}
\def\@codex#1{{\normalfont\ttfamily\hyphenchar\font=-1 #1}\egroup}

\newcommand\pkg{\bgroup\@makeother\_\@makeother\~\@makeother\$\@makeother\^\@codex}
\def\@codex#1{{\normalfont\ttfamily\hyphenchar\font=-1 #1}\egroup}
\makeatother

\newcommand{\unif}{\text{Unif}}

\newcommand{\csp}[1]{\mathcal{#1}}
\newcommand{\poisson}{\text{Poisson}}

\renewcommand{\P}{\mathbb{P}}

\newcommand{\D}{\displaystyle}

\newcommand{\N}{\mathrm{N}}

\newcommand{\E}{\mathbb{E}}

\newcommand{\cov}{\mathrm{cov}}

\newcommand{\indic}{\mathbb{I}}

\begin{document}

\title{Continuous Inference for Aggregated Point Process Data}
\author{Benjamin M. Taylor (Lancaster University), Ricardo Andrade-Pacheco\\ Hugh J. W. Sturrock, (University of California, San Francisco)}
\maketitle

\begin{abstract}
    This article introduces new methods for inference with count data registered on a set of aggregation units. Such data are omnipresent in epidemiology due to confidentiality issues: it is much more common to know the county in which an individual resides, say, than know their exact location in space. Inference for aggregated data has traditionally made use of models for discrete spatial variation, for example conditional autoregressive models (CAR). We argue that such discrete models can be improved from both a scientific and inferential perspective by using spatiotemporally continuous models to directly model the aggregated counts. We introduce methods for delivering (limiting) continuous inference with spatitemporal aggregated count data in which the aggregation units might change over time and are subject to uncertainty. We illustrate our methods using two examples: from epidemiology, spatial prediction malaria incidence in Namibia; and from politics, forecasting voting under the proposed changes to parlimentary boundaries in the United Kingdom.
\end{abstract}

\section{Introduction}

In this article, we will use the phrase `aggregated point process data' (or `aggregated data', for short) to refer to discretely observed data which in reality most likely arose from an underlying spatially- or spatiotemporally-continuous (point) process. A common example of aggregated data, at least in the field of epidemiology, are disease counts observed over a set of regions in space, such as the number of cancer cases in each of the counties of a US state. In this article, we will use the term `agregated models' to refer to statistical models for discretely observed data that are derived from an underlying spatially continuous model; we will see that aggregated models require some sub-region level information which in practice is often available.

Discrete models for aggregated data abound \citep{besag1974,besag1995,rue2009,lee2016}. One possible explanation for this is because point-level information is often not available for economic or confidentiality reasons; another possible explanation is convenience, for example if the desired data are collected routinely alongside other information \citep{beale2010,diggle2010}. To be absolutely clear about our distinction between discrete and aggregated data, we note that it is possible to fit a discrete model and obtain spatially- or spatiotemporally-continuous inference via spatial prediction. For example, we can obtain continuous inference (within regions) from the following model:
\begin{eqnarray}
    T_i &\sim& \poisson{(R_i)}, \label{eqn:discrete}\\
    \log R_i &=& Z_i\beta + Y_i. \nonumber
\end{eqnarray}
Here $T_i$ are the observed events in region $i$ (say $i = 1,\ldots,n$), $Z_i$ are a set of region-specific covariates and $Y_i$ is the value of a continuous spatial process $\csp{Y}$ (e.g. a spatial Gaussian process) at some point within region $i$, such as a population-weighted centroid. The Bayesian paradigm advocates inference via the posterior density, $\pi(\beta,Y_{1:n},\eta|T_{1:n})$, where $\eta$ are parameters describing the properties of the process $\csp{Y}$, e.g. marginal variance and spatial correlation. From this model we can predict the process $\csp{Y}$ anywhere on the plane, so we can obtain a prediction of risk for any point in space at which appropriate values of $Z$ can be found. We regard such models as being spatially discrete because they do not attempt to model the process at the sub-aggregate level.

A more common approach for making inferences for aggregated Poisson counts is to use a spatially discrete model (e.g. a conditional or simultaneous autoregressive or a Besag-York-Mollie model). While the choice of a discrete model with CAR structure may be advantageous in terms of software availability and to some extent, computational efficiency, this type of modelling approach does have some drawbacks:

\begin{enumerate}
    \item[(i)] the notion of `neighbours', upon which the dependence, or lack of dependence between regions in a CAR model is defined is somewhat contrived: neither the size nor shape of regions, nor their internal composition (in terms of population characteristics and location) is taken into consideration in the modelling. A second point is: how should we define neighbours when there are overlaps between regions, or uncertainty in the precise boundary?

    \item[(ii)] for regions that vary widely in shape and size, the covariance structures implied by CAR/SAR models can have counter-intuitive and unappealing properties, see \cite{wall2004};

    \item[(iii)] the modelling assumption of jumps in rates between neighbouring regions is unnatural -- in most cases we would expect risk to vary continuously in space; and

    \item[(iv)] one has to be particularly careful in interpreting aggregated models in order to avoid ecological bias \citep{wakefield2010,wakefield2010a}. For example: individuals living in a region with a large frailty term will not all have the same estimated excess risk attributed to the region.
\end{enumerate}

For certain types of data, aggregation need not be as limiting a factor as it may first seem. The kind of data for which the method proposed in the present article can be applied are those in which the locations of the individual cases (were they known) could be considered as a realisation of a spatiotemoporal Poisson point process \cite{moller2004,baddeley2015}. In the present article, we extend data-augmentation methods developed for aggregated spatial log-Gaussian Cox processes \citep{li2012,taylor2015} in order to deliver continuous inference for aggregated spatiotemporal log-Gaussian Cox processes with overlapping and uncertain boundaries. We do this by incorporating into our model for the aggregated counts additional population and covariate data at the sub-regional level. Such population data are often available e.g. from the Gridded Population of the World \cite{CIESIN2014}.

If we knew the exact location of cases at the sub-aggregation level, then we would be able to fit a spatially continuous point process model e.g. \cite{taylor2015}. Our proposed method employs a data augmentation step for generating putative case locations at the sub-aggregation level using multinomial sampling (making use of additional sub-aggregation-level information) whence the parameters of the continuous model can be updated conditional on the new putative locations. Iterating these steps yields a Gibbs sampler that generates both model parameters for the continuous model as well as realisations of the cases: which can be used to transfer inference from one discretisation to another, among other things. The main novelty in the present article is to extend the research of \citep{li2012,taylor2015} to include the case of spatiotemoporal aggregated data where regional boundaries change over time and may be uncertain. We show how this can be achieved with a simple modfication of the multinomial sampling step. Our modelling approach offers an elegant solution to the issues detailed in (i)--(iv) above.

The methodology discussed in the present article is related to the concept of downscaling \citep{berrocal2010} which, rather than being concerned with the modelling of event counts, is concerned with combining (continuous) information measured at multiple scales, such as that measured by (point-referenced) monitoring locations in continuous space and outputs from numerical models of pollution, say \citep{fuentes2005,matisziw2008}.

Regarding the issue of ecological bias (mentioned above in respect to the disadvantages of discrete models), this occurs when we try to interpret parameter effects at one hierarchy as though they belonged to a different hierarchy, for example we try to interpret aggregate effects as individual-level effects. The aggregate modelling framework we propose in the present article does not completely avoid ecological bias, rather, it enables us to incorporate into our model information at the sub-aggregate level, and thereby come closer to the level of interpretation that might achieved under a truly continuous model.

This article is organised as follows. In Section \ref{sect:basic}, we introduce notation and discuss the simplest case of aggregated spatial data, where there is no overlap between regions and no uncertainty regarding boundaries. We then extend these methods to overlapping (Section  \ref{sect:overlapping}), uncertain (Section \ref{sect:uncertain}) and spatiotemporal datasets (Section \ref{sect:spacetime}). In Section \ref{sect:overlap_assess}, we assess the effect of the amount of overlap on inferences in a spatial dataset for which we know the true location of the cases. Next, in Sections \ref{sect:namibia} and \ref{sect:voting}, we apply our methods to real datasets: respectively modelling spatiotemporal incidence of malaria in Namibia and modelling voting patterns in the last two UK general elections with the aim of predicting what will happen under the proposed boundary changes due to take effect in the 2020 general election. Our article closes with a discussion in Section \ref{sect:discussion}.

\section{Aggregated Spatial Models with Trivial or Empty Intersection of Regions\label{sect:basic}}

As stated above, we assume the existence of some underlying continuous point process $X$ that is responsible for the true, but unobserved pattern of points on the observation window of interest. More specifically, in this article we will assume a Cox model for the intensity function, $R$, of this process, writing the log-intensity for respectively spatial and spatiotemporal processes as:
\begin{equation*}
    \log R(s) = \log P(s) + Z(s)\beta + \csp{Y}(s) \qquad \text{and} \qquad \log R(s,t) = \log P(s,t) + Z(s,t)\beta + \csp{Y}(s,t).
\end{equation*}
Where $P(s)$ (or $P(s,t)$) is a known multiplicative offset for the intensity function, $Z(s)$ (or $Z(s,t)$) is a vector of covariates measured at each location in space (space-time), $\beta$ are unknown parameter effects (to be estimated) and $\csp{Y}$ is a spatial/spatiotemporal process. In practice it is often convenient to assume $\csp{Y}$ is a Gaussian process (yielding the log-Gaussian Cox process (LGCP)) -- and we will do just that in the present article, but the reader should be aware that the models and methods we describe are extensible to a non-Gaussian $\csp{Y}$ too. We will assume that the properties of the process $\csp{Y}$ are parameterised by a vector $\eta$. We will regard this continuous model as our `true' model for the aggregated case counts.

As an example of the techniques we use in this article, we begin with the spatial case, by assuming the discete regions, $A_1,\ldots,A_m$, have at most trivial (line or point) or empty intersection. Inference in this case has previously been described in \cite{taylor2015,diggle2013,li2012}. In this case, our continuous process $X$ is not directly observed, that is we do not observe the exact locations of our points, rather we observe the total number of cases, $T_i$, in region $i$, $A_i$.

In order to perform inference for this model, we aim to produce samples from
\begin{equation}\label{eqn:salgcp}
    \pi(\beta,\eta,Y|T_{1:m}).
\end{equation}
To do this, we use the technique of data augmentation \citep{vandyk2001}. Using a fine grid on which to represent our continuous spatial model (with $Y$ a piecewise-constant representation of the process $\csp{Y}$ on the grid), we introduce parameters $N$ representing the number of cases falling into each grid cell, which is unknown. Instead of sampling from (\ref{eqn:salgcp}) directly, we sample from
\begin{equation}\label{eqn:salgcp_aug}
    \pi(\beta,\eta,Y,N|T_{1:m});
\end{equation}
noting that (\ref{eqn:salgcp}) is a marginal of this density. Sampling from \ref{eqn:salgcp_aug} can be achieved using a Gibbs scheme, alternately drawing from:
\begin{equation*}
    \pi(\beta,\eta,Y|N,T_{1:m}) \qquad \text{followed by} \qquad \pi(N|\beta,\eta,Y,T_{1:m}).
\end{equation*}
If we know $N$, then $T$ is implied, so $\pi(\beta,\eta,Y|N,T_{1:m})=\pi(\beta,\eta,Y|N)$. We can use any sampling method we like for the continuous version of the model to draw $\beta$, $\eta$ and $Y$ from this density. The density $\pi(N|\beta,\eta,Y,T_{1:m})$ is a multinomial density, and so is straightforward to sample from. In practice, it is not necessary to strictly alternate the sampling between these two densities and, depending on the efficiency of the former sampler, it may pay to take several steps in order to increase independence between the successive values of $\beta$, $\eta$ and $Y$ used to perform the multinomial sampling.

One of the subtleties of this method is the need to compute all intersections between the regions and the computational grid on which inference for the continuous surface takes place. This allows us to correctly assign cases to grid cells in the multinomial step of the algorithm: as well as identifying which regions intersect with at particular grid cell, it is also necessary to calculate the area of intersection, the probability a case from region $i$ belongs to cell $j$, $C_j$, is
\begin{equation}\label{eqn:multinom}
    p_{ij} = |A_i \cap C_j|\exp\{Z_j\beta^{(current)} + Y_j^{(current)}\};
\end{equation}
where $Z_j$ is the covariate and value for cell $j$ and $\beta^{(current)}$ and $Y_j^{(current)}$ are respectively the current value of the parameters $\beta$ and $Y_j$ from the MCMC chain used to sample those parameters, the latter being the spatial effect for cell $j$. Note we use the same discretisation for the covariates as we do for the spatial process.

In general these computations incur $O(nm)$ operations, where $n$ is the number of grid cells and $m$ is the number of regions. Thankfully it is only necessary to perform these computations once: for repeated analyses on the same grid and regions the information can be reused.

In \cite{taylor2015}, the authors compare continuous and aggregated inference from a dataset concerning cases of primary biliary cirrhosis in the Newcastle-Upon-Tyne area in the UK. They show that both models produce empirically equivalent inference, albeit the prediction surfaces, as expected, are smoother for the aggregated model.

\section{Aggregated Spatial Models with Non-Trivial Intersection of Regions\label{sect:overlapping}}

For datasets where the set of regions are overlapping non-trivially (that is $\exists i, j$ such that $|A_i\cap A_j|\neq\emptyset$) the sampling procedure described in Section \ref{sect:basic} needs to be modified. These modifications concern the multinomial step of the Gibbs update.

In epidemiology, the most obvious case of overlapping regions occur when data are collected at the health facility level. In the past this has been treated at worst as a point and at best as set of discrete non-overlapping regions. In reality, there will be considerable overlap in regions, arising from health care preferences, availability or whimsically. This issue is still present for aggregated facility level data, i.e. aggregating to districts: facilities that lie near the border of a district will be aggregated to the region they lie in, when in fact cases could have come from across the border. Another example is survey data: cross sectional surveys have traditionally been treated as points (e.g. school surveys for worms/malaria) but in truth, these surveys come might from a school catchment, for example, which often overlap.

\begin{figure}
    \begin{minipage}{0.5\textwidth}
        \begin{center}
            \includegraphics[width=0.9\textwidth]{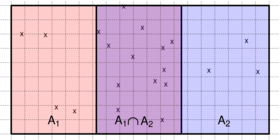}
        \end{center}
    \end{minipage}\begin{minipage}{0.5\textwidth}
        \begin{center}
            \includegraphics[width=0.9\textwidth]{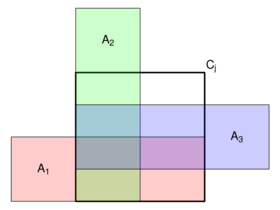}
        \end{center}
    \end{minipage}

    \begin{minipage}{0.5\textwidth}
        \begin{center}
            (a)
        \end{center}
    \end{minipage}\begin{minipage}{0.5\textwidth}
        \begin{center}
            (b)
        \end{center}
    \end{minipage}

    \caption{\label{fig:overlap} Illustrating features that need to be taken into consideration for datasets where region overlapping is non-trivial, see text for details.}
\end{figure}

Suppose we observe $o_i$ events falling into region $i$ i.e. $T_i = o_i$. The multinomial sampling step described in Section \ref{sect:basic} (specifically Equation \ref{eqn:multinom}) assigns each of these events independently to cells $C_j$ intersecting region $i$ with probability proportional to $p_{ij}$. When regions overlap, this would lead to the incorrect number of events being assigned to cells. Consider Figure \ref{fig:overlap} (a), this shows two overlapping regions within each of which the intensity of the process is uniform, when the regions are overlapped, the intensity in $A_1\cap A_2$ is greater than in either $A_1 \backslash (A_1\cap A_2)$ or $A_2 \backslash (A_1\cap A_2)$. Another way of looking at this is that if the underlying $p_{ij}$ was constant over space for each of the small grid cells in Figure \ref{fig:overlap} (a), then the multinomial step described in the section above would not correctly allocate events to grid cells, the result would be inhomogeneous, rather than homogeneous.

Therefore, if we wish to achieve a desired intensity in the overlap region, then we need to correspondingly reduce $p_{ij}$ for cells $j$ that intersect with any intersection between two or more regions, which allows the counts in cells to be contributed from any region involved in the intersection. Figure \ref{fig:overlap} (b) shows the types of of intersections we need to account for when adjusting these probabilities. In this figure, three regions overlap a grid cell, but of course in a real-life dataset there may be many more overlapping regions and these may intersect non-trivially, for example $A_{i_1}\cap A_{i_2}\cap C_j$ may not be a connected set.

In Figure \ref{fig:overlap} (b) there are $\sum_{i=0}^3 {3 \choose i} - 1 = 7$ different intersections to take acount of (not including the area of $C_j$ that does not belong to any $A_i$) and in general, if $r$ regions overlap a cell, there will be $O(2^r)$ intersections to account for. A further complication is that if (on the unobserved continuum) an event falls in the intersection of $r'$ regions i.e. for some $x\in X\cap C_j$, we have $x\in \bigcap_{k=1}^{r'} A_{i_k}$, then $x$ may not be assigned to the aggregate total for each of $A_{i_1},\ldots,A_{i_{r'}}$ with equal probability i.e. the condition
\begin{equation*}
    \P\left[x\in A_{i_1} \big| x\in \bigcap_{k=1}^{r'} A_{i_k}\right] = \cdots = \P\left[x\in A_{i_{r'}} \big| x\in \bigcap_{k=1}^{r'} A_{i_k}\right]
\end{equation*}
may not be true. In practice, one might assume that each region $A_i$ has an associated measure of `sampling effort', $e_i$ so that for example the allocation of points falling in an intersection $A_1\cap A_2$ can be assigned to region $A_1$ region with probability $e_1 / (e_1 + e_2)$. Two further alternatives are as follows: firstly, we could also allow the sampling effort for region $i$ to vary between computational cells $C_j$ that intersect that region, leading to cell-specific allocation probabilities $e_{1j} / (e_{1j} + e_{2j})$; secondly, we could allow sampling effort to vary between all multi-way intersections of regions (see below), leading to an allocation probability of $e_{1jk} / (e_{1jk} + e_{2jk})$ for each intersection. However, in real life applications we suspect these last two quantities, and in particular the latter, may be difficult to ascertain.

In the case of two regions $A_1$ and $A_2$ intersecting a grid cell $C_j$, the correct modification to Equation \ref{eqn:multinom} for region $A_1$ would look like:

\begin{eqnarray*}
    p_{1j}^{\text{mod}} &=& (|A_1 \cap C_j| - |A_1 \cap A_2 \cap C_j|)\exp\{Z_j\beta^{(current)} + Y_j^{(current)}\} + \\
    & & \qquad\qquad \P(x\in A_1 | x \in A_1 \cap A_2 \cap C_j)|A_1 \cap A_2 \cap C_j|\exp\{Z_j\beta^{(current)} + Y_j^{(current)}\}\\
    &=& p_{1j}\times\left\{\left(1-\frac{|A_1 \cap A_2 \cap C_j|}{|A_1 \cap C_j|}\right) + \P[x\in A_1 | x\in A_1 \cap A_2 \cap C_j]\frac{|A_1 \cap A_2 \cap C_j|}{|A_1 \cap C_j|}\right\} \\
    &=& p_{1j}\times\{\underbrace{\P[x\in A_1 | x\in(A_1 \cap C_j)\backslash(A_1 \cap A_2 \cap C_j)]}_{=1}\P[x\in(A_1 \cap C_j)\backslash(A_1 \cap A_2 \cap C_j)] +  \\
    & & \qquad\qquad\qquad\qquad\qquad\qquad\qquad  \P[x\in A_1 | x\in A_1 \cap A_2 \cap C_j]\P[x\in A_1 \cap A_2 \cap C_j]\} \\
    &=& p_{1j}q_{1j}
\end{eqnarray*}
where $p_{1j}$ is as defined in Equation \ref{eqn:multinom}. In general for a grid cell in which $r$ regions potentially intersect, the scaling factor for $p_{1j}$ (i.e. the term $q_{1j}$) would become rather complex; fortunately this term can be seen as an expectation, which can be evaluated approximately using Monte Carlo methods. Let $\Pi^{(j)}=\{\Omega_1^{(j)},\ldots,\Omega_{2^{r_j}}^{(j)}\}$ denote the partition of cell $C_j$ into disjoint subsets $\Omega_1^{(j)},\ldots,\Omega_{2^{r_j}}^{(j)}$ i.e. each element of $\Pi^{(j)}$ is one of the $2^{r_j}$ possible multi-way intersections of the regions $A_{i_1},\ldots,A_{i_{r_j}}$ overlapping cell $C_j$ discussed above and thus $C_j=\bigcup\limits_k^{\cdot}\Omega_k^{(j)}$.

Examining the structure of $q_{1j}$, it can be seen that for a general region $i$ intersecting cell $j$ we will have:
\begin{equation}\label{eqn:modif}
    q_{ij} = \sum_{k=1}^{2^{r_j}}\P(x\in A_i|x\in\Omega_k^{(j)})\P(x\in\Omega_k^{(j)})
\end{equation}
since $\P(x\in A_i|x\in\Omega_k^{(j)})=0$ for $\Omega_k^{(j)}\cap A_i=\emptyset$. We can write this as an expectation taken with respect to a uniformly ditributed random variable on $C_j$:
\begin{eqnarray*}
    q_{ij} &=& \E\left[\P(x\in A_i|x\in\Omega_k^{(j)})\indic(x\in\Omega_k^{(j)})\right]\\
    &=& \E\left[W_{ijk}\indic(x\in\Omega_k^{(j)})\right]
\end{eqnarray*}
Another way to view $q_{ij}$ is as $\P(\text{$x$ contibutes to the aggregate total of region $A_i$ in cell $C_j$})$.

If sampling effort is constant across space, then
\begin{equation*}
    W_{ijk} = W_{ij} = \frac{e_i}{\D\sum_{l : A_i \cap \Omega_l^{(j)}\neq\emptyset}e_l},
\end{equation*}
and we can estimate $q_{ij}$ by drawing iid $u_1,\ldots,u_M\sim\unif(C_j)$ and using
\begin{equation*}
    \hat q_{ij} = \frac1M\sum_{l=1}^M\left[\frac{e_i}{\sum_{v : A_i \cap \Omega_v^{(j)}\neq\emptyset}e_v}\indic(u_l\in\Omega_k^{(j)})\right]
\end{equation*}
This method is implemented as an extension to the \proglang{R} package \pkg{lgcp}: the main assumption is that the $W_{ijk}$ are known; if this is not the case, a pragmatic (but informative) assumption might be $W_{i,j,k}=c$ whence events are allocated at random i.e. without preference for a particular $A_i$.

\begin{algorithm}
    \caption{Generic Algorithm For Fixed Regions, including overlapping and spatiotemporal data.}
    \label{alg:generic_non_stochastic}
    \begin{algorithmic}[1]
        \STATE Project model covariates and population off set onto the inferential grid.
        \STATE Compute and store information on overlaps between inferential grid and (possibly overlapping) polygons. This includes which polygons intersect each grid cell and the area of the intersection.

        \STATE Initialise $N$ on the computational grid. In the absence of other information this could be achieved by sampling $N$ proportional to the offset, respecting regional boundaries; or if no offset is available, then uniformly within each region.
        \STATE Initialise the sampling algorithm for $\pi(\beta,\eta,Y|N,T_{1:m})=\pi(\beta,\eta,Y|N)$.
        \FOR{$i=1,\ldots,n$, where $n$ is large}
            \STATE Update $\beta$, $\eta$, $Y$ conditional on $N$ (e.g. using MCMC).
            \IF{$i\mod r_f = 0$, where $r_f$ is the resampling frequency for $N$}
                \STATE Update $N$ conditional on $T$ and current values of $\beta$, $\eta$, $Y$ using multinomial sampling.
            \ENDIF
        \ENDFOR
        \STATE Result is a set of samples from $\pi(\beta,\eta,Y,N|T)$
    \end{algorithmic}
\end{algorithm}

\section{Aggregated Spatial Models with Uncertainty in Regional Boundaries\label{sect:uncertain}}

Our discussion in Sections \ref{sect:basic} and \ref{sect:overlapping} has thus far focussed on the case that each region $A_i$ is a known fixed area in space. However, in practical problems, there can be uncertainty in catchment boundaries. For example, while health facility catchment area boundaries can be defined by governments or via models, individuals may not seek treatment at their nearest facility \citep{noor2006,guagliardo2004,tanser2001,akin1999}. Equally, health facility catchments may span more than one district. Aggregated case numbers may therefore include cases who acquired their infection in a neighboring district. Note that as in Section \ref{sect:overlapping}, these districts may be overlapping.

To account for uncertainty in areas $\{A_i\}$ we replace these fixed regions by stochasic regions $\{\Phi_i(\Gamma_i)\}$ the physical size, shape and location of which is controlled by a random variable $\Gamma_i$. There are many choices for the statistical distribution of $\Gamma_i$, $\pi(\Gamma_i)$, and in practice, the way we choose it this will be context dependent (on the spatial sampling strategy we are trying to describe). It could be that we wish to impose different parametrisations on each region, and we may even wish to treat some regions as fixed. Here are two examples
\begin{enumerate}
    \item $\Phi_i(\Gamma_i)$ could be a contraction or expansion by a factor $\Gamma_i$ of some fixed polygon towards or from the centroid (or some other appropriate point). $\Gamma_i$ in this case would be a positive-valued continuous random variable.
    \item $\Phi_i(\Gamma_i)=A_{il}$ could be selected at random from a set of fixed polygons (or a combination of points and polygons) $\{A_{i1},\ldots,A_{iN_i}\}$ with weight proportional to $\{\gamma_{i1},\ldots,\gamma_{iN_i}\}$. In other words $\Gamma_i$ would be a discrete random variable describing the mixture index and taking values in $\{1,\ldots,N_i\}$ with $A_{il}$ being selected with probability $\gamma_{il}/(\gamma_{i1}+\cdots+\gamma_{iN_i})$.
\end{enumerate}

In a similar way to how we modified Equation \ref{eqn:multinom} to account for overlapping regions, we can use the same technique in the case of uncertain boundaries. Each realisation of parameters $\gamma=(\gamma_1,\ldots,\gamma_m)$ drawn from the joint distribution of $\Gamma=(\Gamma_1,\ldots,\Gamma_m)$, $\pi(\Gamma)$, (we permit correlation in the $\Gamma_i$s) gives rise to a different partition of space into regions $A_{\gamma,1},\ldots,A_{\gamma,m}$ and a different set of intersections for each cell, $\Pi_\gamma^{(j)}=\{\Omega_{\gamma,1}^{(j)},\ldots,\Omega_{\gamma,2^{r_j}}^{(j)}\}$. For this choice of parameters $\gamma$, Equation \ref{eqn:modif} holds with a minor extension of notation as detailed in the present section, but in order to calculate an $q_{ij}$ we now need to take account of the uncertainty in $\gamma$. Let $\gamma^{(l)}$ denote the $l$th sample of $n_\gamma$ realisations $\gamma\sim\pi(\Gamma)$. we propose using the following

\begin{equation*}
    \hat q_{ij}^{\text{unc}} = \frac1{Mn_\gamma}\sum_{g=1}^{n_\gamma}\sum_{l=1}^M\left[\frac{e_i}{\sum_{v : A_{\gamma,i} \cap \Omega_{\gamma,v}^{(j)}\neq\emptyset}e_v}\indic(u_l\in\Omega_{\gamma,k}^{(j)})\right]
\end{equation*}
Note that we are assuming the hyperparameters of the joint density of $\Gamma$ are known. While of course it would, in theory, be possible to learn these parameters by adding an additional hierarchy into our model, doing so would affect our ability to identify correlation in the Gaussian field, an already very challenging inferential problem \cite{zhang2004}. We are also assuming that $\Gamma$ is independent from $Y$, $\beta$, $\eta$.

\begin{algorithm}
    \caption{Generic Algorithm For Stochastic Regions, including overlapping and spatiotemporal data. Note that in this algorithm the cost of finding overlaps and intersections with the inferential grid is potentially high. If the reader has access to parallel processing facilities, then the part of the algorithm that samples $\beta$, $\eta$, $Y$ can run independently from the part that involves computation of overlaps and intersections. If overlaps and intersections can be computed in the same or less CPU time as it takes for $r_f$ iterations of the sampling scheme for $\pi(\beta,\eta,Y|N)$, then there is no loss in performance compared with Algorithm \ref{alg:generic_non_stochastic}.}
    \label{alg:generic_stochastic}
        \begin{algorithmic}[1]
        \STATE Project model covariates and population offset onto the inferential grid.
        \STATE Simulate $\gamma\sim\pi(\Gamma)$ and compute partitions $A_{\gamma,1},\ldots,A_{\gamma,m}$ and overlaps between inferential grid and these (possibly overlapping) polygons, including areas of each intersection.
        \STATE Initialise $N$ given the initial choice of $A_{\gamma,1},\ldots,A_{\gamma,m}$ above as per Algorithm \ref{alg:generic_non_stochastic}
        \STATE Initialise the sampling algorithm for $\pi(\beta,\eta,Y|N,T_{1:m})=\pi(\beta,\eta,Y|N)$.
        \FOR{$i=1,\ldots,n$, where $n$ is large}
            \STATE Update $\beta$, $\eta$, $Y$ conditional on $N$ (e.g. using MCMC).
            \IF{$i\mod r_f = 0$, where $r_f$ is the resampling frequency for $N$}
                \STATE Simulate $\gamma\sim\pi(\Gamma)$, compute partitions $A_{\gamma,1},\ldots,A_{\gamma,m}$ and overlaps.
                \STATE Update $N$ conditional on $T$ and current values of $\beta$, $\eta$, $Y$ using multinomial sampling (ensuring to respect new boundaries from the previous step).
            \ENDIF
        \ENDFOR
        \STATE Result is a set of samples from $\pi(\beta,\eta,Y,N|T)$
    \end{algorithmic}
\end{algorithm}

\section{Aggregated Spatio-Temporal Models\label{sect:spacetime}}

Aggregated data are often captured at the regional level over a period of time and when this is the case, it is usually of scientific interest to model the process in both space and time. A common practical issue with aggregated data collected over a period of time is that it is often the case that regional boundaries change, or that regions merge, or split apart. Over time, we could also encounter any combination of the types of overlapping discussed in Sections \ref{sect:basic}--\ref{sect:uncertain}.

Extension of the methods above to the spatiotemporal case is in fact quite straightforward in principle, although with the addition of the temporal dimension there will obviously be a greater computational burden. We require a continuous model for the spatiotemporal process,
\begin{eqnarray*}
    N(s,t) &\sim& \text{Poisson}(R(s,t)),\\
    \log R(s,t) &=& P(s,t) + Z(s,t)\beta + \csp{Y}(s,t),
\end{eqnarray*}
where $N(S,t)$ is the number of  and a method to deiver inference for that process on a fine grid. Such methods are discussed in \cite{brix2001}, \cite{diggle2005} and \cite{taylor2015}, for example. These authors employ a Metropolis-adjusted Langevin algorithm, but more sophisticated sampling schemes, such as the Riemann manifold Hamiltonian Monte Carlo sampling scheme discussed in \cite{girolami2011} can also be extended to deliver inference for the Cox process. The method introduced by \cite{brix2001} advocates analysing a series of spatiotemporal data by splitting them into chunks; the time dimension, as well as the space dimension are discretised. The main idea is to exploit the fact that for most inferential problems we find that not all data in the history of the process will affect the intensity at time $t$: we might only need data between time $t-k$ and $t$ to deliver inference for time $t$. We can therefore provide inference for a series of times $t_1,t_2,\ldots,t_m$ by combining inferences from $[t_1-k,t_1],[t_2-k,t_2],\ldots,[t_m-k,t_m]$.

We will assume that inference proceeds on a fine spatiotemporal grid, in which case the data augmentation scheme described in Section\ref{sect:basic} applies once more: we alternately update $\beta$, $Y$ and $\eta$ conditional on $N$ followed by $N$ conditional on $\beta$, $Y$ and $\eta$. A number of options are available in terms of the continuous model, for cells $(s_1,t_1)$ and $(s_2,t_2)$ on the computational grid (where $s$ represents space and $t$ represents time), two examples could include: a separable spatiotemporal covariance function for $Y$:
\begin{equation*}
   \cov[Y(s_1,t_1),Y(s_2,t_2)] = f_1(||s_2-s_1||;\eta_s)f_2(|t_2-t_1|;\eta_t),
\end{equation*}
or a non-seperable function, which allows for interaction between space and time:
\begin{equation*}
   \cov[Y(s_1,t_1),Y(s_2,t_2)] = f(||s_2-s_1||,|t_2-t_1|;\eta),
\end{equation*}
here assuming a $Y$ is second-order stationary.

\section{How Does The Amount of Overlap Affect Inference?\label{sect:overlap_assess}}

In this section, we use a point process dataset of 415 geo-referenced cases of definite or probable primary biliary cirrhosis (PBC) alive between 1987 and 1994 to illustrate the effect of increasing the amount of overlap between regions on inference. These point-referenced data were originally analysed in \cite{prince2001} and later \cite{taylor2015} compared inferences obtained from an aggregated version of the data to those obtained assuming an effectively continuous risk surface.

\begin{figure}[htbp]
    \begin{minipage}{0.5\textwidth}
        \includegraphics[width=\textwidth]{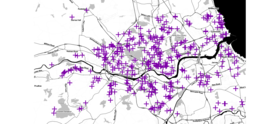}
    \end{minipage}\begin{minipage}{0.5\textwidth}
        \includegraphics[width=1.15\textwidth]{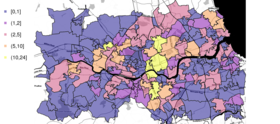}
    \end{minipage}
    \caption{\label{fig:cirrhosis_data}Figure illustrating exact locations of cases (left) and (case counts aggregated to regions).}
\end{figure}

\begin{figure}[htbp]
    \begin{minipage}{0.333\textwidth}
        \includegraphics[width=\textwidth]{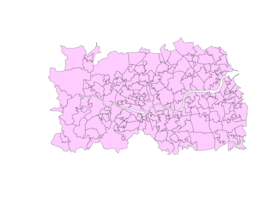}
    \end{minipage}\begin{minipage}{0.333\textwidth}
        \includegraphics[width=\textwidth]{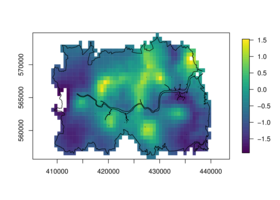}
    \end{minipage}\begin{minipage}{0.333\textwidth}
        \includegraphics[width=\textwidth]{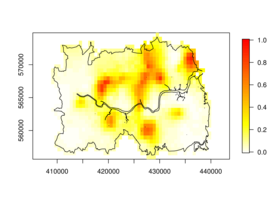}
    \end{minipage}

    \begin{minipage}{0.333\textwidth}
        \includegraphics[width=\textwidth]{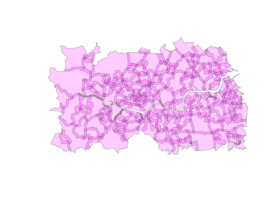}
    \end{minipage}\begin{minipage}{0.333\textwidth}
        \includegraphics[width=\textwidth]{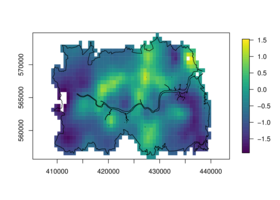}
    \end{minipage}\begin{minipage}{0.333\textwidth}
        \includegraphics[width=\textwidth]{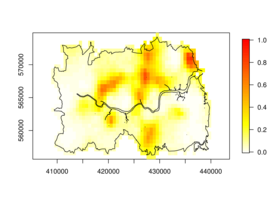}
    \end{minipage}

    \begin{minipage}{0.333\textwidth}
        \includegraphics[width=\textwidth]{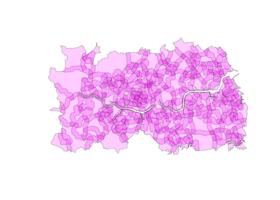}
    \end{minipage}\begin{minipage}{0.333\textwidth}
        \includegraphics[width=\textwidth]{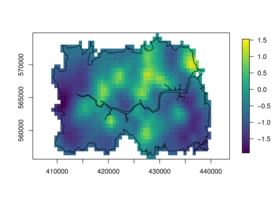}
    \end{minipage}\begin{minipage}{0.333\textwidth}
        \includegraphics[width=\textwidth]{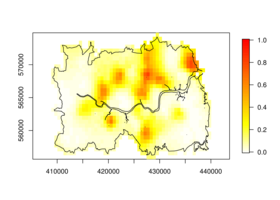}
    \end{minipage}

    \begin{minipage}{0.333\textwidth}
        \includegraphics[width=\textwidth]{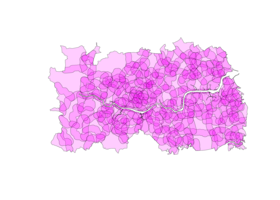}
    \end{minipage}\begin{minipage}{0.333\textwidth}
        \includegraphics[width=\textwidth]{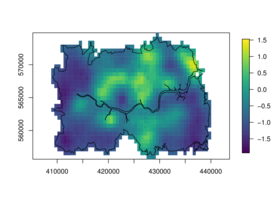}
    \end{minipage}\begin{minipage}{0.333\textwidth}
        \includegraphics[width=\textwidth]{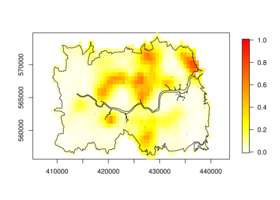}
    \end{minipage}

    \begin{minipage}{0.333\textwidth}
        \includegraphics[width=\textwidth]{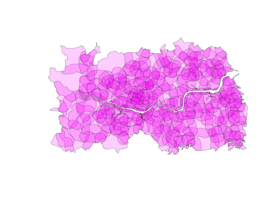}
    \end{minipage}\begin{minipage}{0.333\textwidth}
        \includegraphics[width=\textwidth]{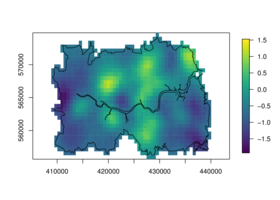}
    \end{minipage}\begin{minipage}{0.333\textwidth}
        \includegraphics[width=\textwidth]{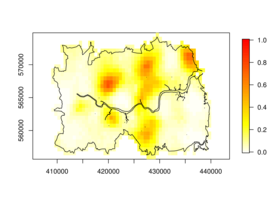}
    \end{minipage}
    \caption{\label{cirrhosis_results} Left: the overlapping regions. Middle: $\E(Y|\text{data})$ i.e. the posterior mean of the latent field, $Y$. Right: $\P(\exp{Y}>2|\text{data})$ i.e. the posterior probability that the relative risk exceeds 2. Top row to bottom row: results from aggregated dataasets incorporating respectively a 0m, 150m, 300m, 450m and 600m overlap.}
\end{figure}

The left plot in Figure \ref{fig:cirrhosis_data} shows the locations of the 415 cases and the right hand plot shows the total number of cases in each of the aggregation units. We used a subset of the indices of deprivation originally measured at the Lower Super Output Area (LSOA) level as covariates in our Poisson regression model: Income, Barriers, Environment and also included an intercept term. We only used a subset of the IMD domains because there was some collinearity between the domains measured in this region: where two variables were correlated, we removed the domain least likely to explain the incidence. We rasterised the covariates onto the grid shown in the middle and right-hand plots in Figure \ref{cirrhosis_results}. We included a population offset from data taken at the LSOA level and rasterised as per the covariates. We created five datasets based on the point-locations and the aggregation units illustrated in the left column of Figure \ref{cirrhosis_results}.

The five different aggregation units were created by enlarging each of the individual regions using buffers of size 0 (giving the data illustrated in the righ hand plot of  Figure \ref{fig:cirrhosis_data}), 150, 300, 450 and 600 metres and cropping the resulting regions so the boundary was identical to that in Figure \ref{fig:cirrhosis_data}. Each point that fell into a non-trivial intersection of regions was randomly allocated to one of the regions. We used an exponential model for the stationary spatial covariance function, so that two points a distance $d$ apart would have the following covariance: $\sigma^2\exp\{-d / \phi\}$.

Full details of the MCMC scheme employed in the analysis are given in \cite{taylor2015}. We ran the MCMC chains for each dataset for 500,000 iterations with a 10,000 iteration burnin and retaining every 490th sample. For the covariates, we used zero-mean independent Gaussian priors with standard deviation 1000, the log standard deviation of the marginal variance of the latent field, $\sigma$ had a Gaussian prior with mean 0 and standard deviation 0.3 and the prior for the spatial dependence parameter, $\phi$ had mean $\log(1500)$ and standard deviation $0.09$; the latter was chosen so as to avoid numerical singularity in the discrete Fourier transform employed to handle matrix operations (see \cite{taylor2015} for further detail) and gives a range of up to approximately 1/5 of the height of the observation window.

\begin{table}[htbp]
    \tiny
    \centering
    \begin{tabular}{l|l|l|l|l|l}
        Parameter & 0m & 150m & 300m & 450m & 600m \\ \hline
        $\beta_{\text{Intercept}}$ & -17.4(-18.6, -16.18) & -17.92(-19.22, -16.6) & -18.64(-19.91, -17.31) & -19.22(-20.64, -17.76) & -18.99(-20.44, -17.57) \\
        $\beta_{\text{Income}}$ & 5.03(2.59, 7.3) & 4.08(1.4, 6.89) & 3(0.19, 5.83) & 2.95(0.17, 6.17) & 1.51(-1.53, 4.82) \\
        $\beta_{\text{Barriers}}$ & -0.15(-0.19, -0.1) & -0.14(-0.19, -0.08) & -0.11(-0.17, -0.06) & -0.09(-0.15, -0.03) & -0.11(-0.18, -0.05)\\
        $\beta_{\text{Environment}}$ & 0(-0.05, 0.04) & 0.01(-0.03, 0.05) & 0.02(-0.02, 0.07) & 0.01(-0.04, 0.06) & 0.03(-0.02, 0.08) \\
        $\log(\sigma)$ & 0.07(-0.18, 0.3) & 0.06(-0.22, 0.32) & 0.06(-0.21, 0.31) & 0.08(-0.19, 0.37) & -0.03(-0.32, 0.24) \\
        $\log(\phi)$ & 7.35(7.18, 7.53) & 7.34(7.18, 7.51) & 7.33(7.16, 7.5) & 7.32(7.14, 7.49) & 7.33(7.15, 7.51)
    \end{tabular}
    \caption{\label{tab:cirrhosis_parameters} Parameter estimates from the Cirrhosis example.}
\end{table}

The parameter estimates from each of the datasets is given in Table \ref{tab:cirrhosis_parameters}. Examining this table, it can be seen that the effect of income deprivation on incidence appears to be attenuated in magnitude and level of significance by increasing the amount of overlap with an effect size ranging between 5.03(2.59, 7.3) when there is no overlap to 1.51(-1.53, 4.82) when the overlap is 600m. The effect of changing the overlap on the barriers domain of the IMD was less pronounced with effect sizes being broadly similar, ranging for example from -0.15(-0.19, -0.1) in the no overlap dataset to -0.11(-0.18, -0.05) for the 600m overlap dataset. There was little difference in the poterior estimates of the enviroment deprivation covariate and also in the estimates of$\log(\sigma)$ and $\log(\phi)$. Though examining a plot of the prior and posterior for $\log(\phi)$, it appears that the data  provide little information on this parameter.

Figure \ref{cirrhosis_results} shows how the estimated mean and exceedance probabilities change under these difference levels of aggregation. These plots show similar pattern in the estimated mean and exceedances across the different scenarios, with visual differences being most likely attributable to the specific allocation of points to the aggregation units in each of the scenarios.

\section{Application: Analysis of Health Facility Data in Namibia with Overlapping Catchment Areas \label{sect:namibia}}

\begin{figure}[htbp]
    \centering

    \includegraphics[width=\textwidth]{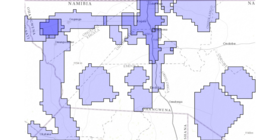}
    \caption{\label{fig:catch} Showing the 16 Catchment Areas surrounding the following clinics: Edundja Clinic, Endola Clinic, Engela District Hospital, Eudafano Clinic, Odibo Health Centre, Ohalushu Clinic, Ohangwena Clinic, Ohaukelo Clinic, Okambebe Clinic, Okatope Clinic, Omungwelume Clinic, Onamukulo Clinic, Ondobe Clinic, Onekwaya Clinic, Ongenga Clinic and Ongha Health Centre.}
\end{figure}

In this section, we analyse a spatiotemporal dataset from northern Namibia. These data consist of monthly counts of malaria cases over a 2 year period reporting to one of 16 clinics/hospitals in the study area, shown in Figure \ref{fig:catch}. Malaria incidence in this area is at the pre-elimination level, with approximately 5.8 cases reported per month.

\begin{figure}[htbp]
    \centering

    \includegraphics[width=0.5\textwidth]{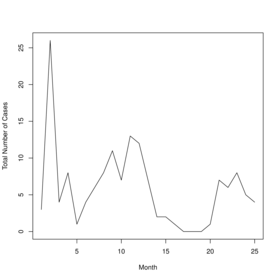}
    \caption{\label{casespermonth} Number of cases reported by month}
\end{figure}

Figure \ref{casespermonth} shows the total number of cases of malaria reported to each of the health facilities over each month of the study period. We fitted a spatiotemoporal log-Gaussian Cox process to the monthly case counts. The grid-level version of this model took the form:

\begin{eqnarray}\label{eqn:stppmodel}
   X_{it} &=& \text{Poisson}[R_{it}]\\
   R_{it} &=& C_A\lambda_{it}\exp\{Z_{it}\beta+Y_{it}\}\nonumber
\end{eqnarray}
In the above model, $X_{it}$ is the total number of events in the cell indexed $i$ of the computational grid at time index $t$, $R_{it}$ is the rate of the Poisson process, $C_A$ is the cell area, $\lambda_{it}$ is a known offset, $Z_{it}$ is a vector of measured covariates (see below) and $Y_{it}$ is the value of the latent Gaussian process at the centroid of computational grid cell $(i,t)$.

We assumed a separable covariance function for the dependence structure of $Y$, induced through a transformation of a sequence of serially-correlated random variables, $\{\Gamma_1,\ldots,\Gamma_T\}$ where $T$ is the maximum number of time points and $\Gamma_k = (\Gamma_{1k},\ldots,\Gamma_{Mk})^T$ is a column vector. We used the transformation $Y_t = -\frac{\sigma^2}{2} + \Sigma^{1/2}_{\sigma,\phi}\Gamma_{t}$, where $\sigma$ is the marginal standard deviation of $Y$ and $\phi$ is the spatial dependence parameter, for which, we assumed an exponential correlation function. The matrix $\Sigma^{1/2}_{\sigma,\phi}$ was the matrix `square root' of the covariance matrix of the spatial process conditional on $\sigma$ and $\phi$ evaluated at the centroids of the spatial grid, with cells arranged in lexicographic order as in \cite{moller1998}.

The motivation for working with a transformation for $Y$ is because we can put a simple prior on $\{\Gamma_k\}$:
\begin{eqnarray*}
    \pi(\Gamma_1,\ldots,\Gamma_T|\theta) &=& \pi(\Gamma_1)\pi(\Gamma_2|\Gamma_1)\cdots\pi(\Gamma_T|\Gamma_{T-1}),\\
    &\sim& \N(\Gamma_1;0,1)\prod_{t=2}^T \N[a_{\delta_t}(\theta)\Gamma_{t-1},1-a_{\delta_t}(\theta)^2],
\end{eqnarray*}
where $a_{\delta_t}(\theta) = \exp{-\theta(\delta_t)}$ and $\delta_t$ is the time between the $t$th and the $(t-1)$th time index. We completed our model by assuming Gaussian priors for the log of $\sigma$, $\phi$ and $\theta$:
\begin{eqnarray*}
    \log\sigma&\sim&\N(0,0.3^2)\\
    \log\phi&\sim&\N(\log(0.015),0.3^2)\\
    \log\theta&\sim&\N(0,1)
\end{eqnarray*}
and independent $\N(0,100^2)$ priors for each component of $\beta$. These are relatively diffuse priors, with the exception of the prior for $\phi$, which apriori sets the range of spatial dependence to be up to around 1/5 of the width of the observation window: this is required to avoid numerical sigularities in matrix computations, see \cite{taylor2015}. Note that the units of $\phi$ are in latitude/longitude in this example.
% betaprior <- spatsurv::betapriorGauss(mean=0,sd=100)
% omegaprior <- spatsurv::omegapriorGauss(mean=0,sd=1)
% etaprior <- spatsurv::etapriorGauss(mean=log(c(1,0.015)),sd=c(0.3,0.3))

As in \cite{moller1998} and \cite{taylor2015}, we used the fast Fourier transform (FFT) to handle matrix computations (which gives a massive saving in terms of speed and also disk usage) and an adaptive Metropolis Langevin Algorithm (MALA) to draw from $\pi(Y,\eta,\beta|N)$. While these choices above reflect our personal preference for model and MCMC algorithm, partly due to our significant previous experience with MALA; the reader should be reminded that essentially any MCMC algorithm can be used to sample from this density. Our choice has the following advantages:
\begin{itemize}
    \item Ignoring the convergence rate of the MCMC chain, which is partially data-dependent (in this example we achieve convergence and good mixing in 500,000 iterations), the method is $O(T)$ in time, and $O(M\log M)$ in space, where $M$ is the number of spatial computational cells.
    \item Due to not having to store full covariance matrices, the method is $O(M)$ in terms of storing $\Sigma^{1/2}_{\sigma,\phi}$.
\end{itemize}
One disadvantage is the wrap-around effects from the FFT which we can avoid, as in this example, by extending the grid by two in each direction. We used a $128\times64$ computational grid, yielding output on a $64\times32$ grid.

We used the following covariates: monthly enhanced vegetation index (EVI), monthly normalized water difference index NDWI, and monthly land surface temperature (LST), each of which was available on a $44\times76$ raster for each month from the MODIS satellite. Although we had access to it, we did not include elevation as a covariate in our model because it did not vary much over our study region. THe EVI, NDWI and LST variables were obtained via Google Earth Engine. We lagged each of the included covariates by one month for our analysis. Malaria case data occurring between May 2010 to May 2012 were collected retrospectively from patient registers from the 16 health facilities in Engela district, Namibia. Cases were defined as those individuals with a positive malaria diagnosis by rapid diagnostic test (RDT). 2010 population density data for the district were obtained from WorldPop (\url{www.worldpop.org.uk}).

The probability of attendance at each of the 16 facilities was calculated using a gravity Huff model \cite{huff1964}. This means that the probabilities are assumed to be directly proportional to the relative `utility' that each facility represents to a patient. In our case, we defined the travel time to the health facility as the only driver of the utility function. Then the probability of patient $i$ attending facility $j$ can be expressed as
\begin{equation*}
    p_{ij} = \frac{d_{ij}^\delta}{\sum_j d_{ij}^\delta}
\end{equation*}
where $d_ij$ is the travel time from the household of patient $i$ to facility $j$, and $\delta$ is a sensitivity parameter of the model. We created the 16 catchment boundaries by using a cutoff of 20\% on the probabilities $p_{ij}$.

\begin{figure}[htbp]

    \centering

    \begin{minipage}{0.333\textwidth}
        \includegraphics[width=\textwidth]{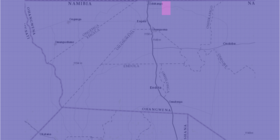}
    \end{minipage}\begin{minipage}{0.333\textwidth}
        \includegraphics[width=\textwidth]{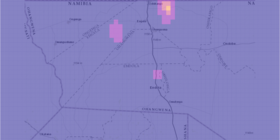}
    \end{minipage}\begin{minipage}{0.333\textwidth}
        \includegraphics[width=\textwidth]{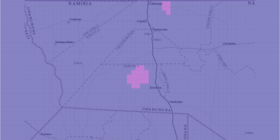}
    \end{minipage}

    \begin{minipage}{0.333\textwidth}
        \centering May 2010
    \end{minipage}\begin{minipage}{0.333\textwidth}
        \centering June 2010
    \end{minipage}\begin{minipage}{0.333\textwidth}
        \centering July 2010
    \end{minipage}

    \begin{minipage}{0.333\textwidth}
        \includegraphics[width=\textwidth]{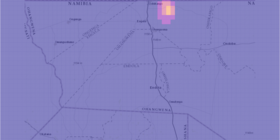}
    \end{minipage}\begin{minipage}{0.333\textwidth}
        \includegraphics[width=\textwidth]{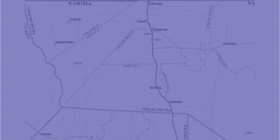}
    \end{minipage}\begin{minipage}{0.333\textwidth}
        \includegraphics[width=\textwidth]{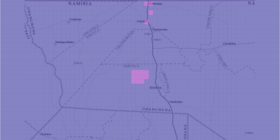}
    \end{minipage}

    \begin{minipage}{0.333\textwidth}
        \centering August 2010
    \end{minipage}\begin{minipage}{0.333\textwidth}
        \centering September 2010
    \end{minipage}\begin{minipage}{0.333\textwidth}
        \centering October 2010
    \end{minipage}

    \begin{minipage}{0.333\textwidth}
        \includegraphics[width=\textwidth]{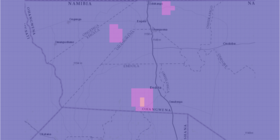}
    \end{minipage}\begin{minipage}{0.333\textwidth}
        \includegraphics[width=\textwidth]{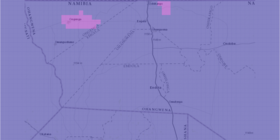}
    \end{minipage}\begin{minipage}{0.333\textwidth}
        \includegraphics[width=\textwidth]{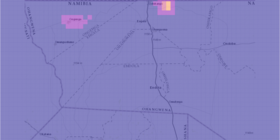}
    \end{minipage}

    \begin{minipage}{0.333\textwidth}
        \centering November 2010
    \end{minipage}\begin{minipage}{0.333\textwidth}
        \centering December 2010
    \end{minipage}\begin{minipage}{0.333\textwidth}
        \centering January 2011
    \end{minipage}

    \begin{minipage}{0.333\textwidth}
        \includegraphics[width=\textwidth]{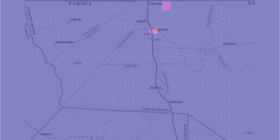}
    \end{minipage}\begin{minipage}{0.333\textwidth}
        \includegraphics[width=\textwidth]{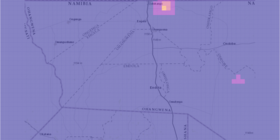}
    \end{minipage}\begin{minipage}{0.333\textwidth}
        \includegraphics[width=\textwidth]{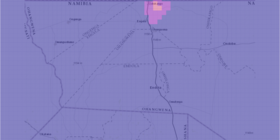}
    \end{minipage}

    \begin{minipage}{0.333\textwidth}
        \centering February 2011
    \end{minipage}\begin{minipage}{0.333\textwidth}
        \centering March 2011
    \end{minipage}\begin{minipage}{0.333\textwidth}
        \centering April 2011
    \end{minipage}
    \caption{\label{fig:months_1_12} Months 1--12. Plots showing $\P[\exp\{Y\}>1.5|\text{data}]$ i.e. the posterior probability that the relative risk is greater than 1.5. Colour key: \protect\includegraphics[width=1em]{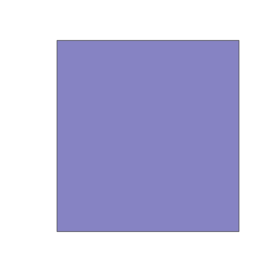} [0.0,0.2], \protect\includegraphics[width=1em]{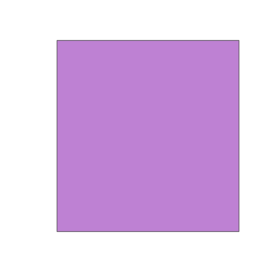} (0.2,0.4], \protect\includegraphics[width=1em]{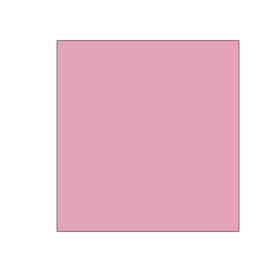} (0.4,0.6], \protect\includegraphics[width=1em]{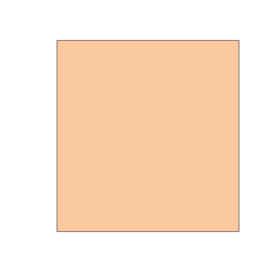} (0.6,0.8], \protect\includegraphics[width=1em]{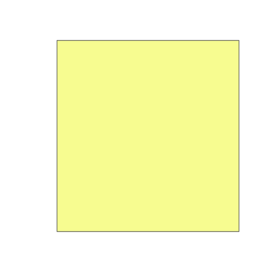} (0.8,1.0].}
\end{figure}

As in the cirrhosis analysis above, we ran the MCMC chain for 500,000 iterations with a 10,000 iteration burnin and retaining every 490th sample for analysis.

\begin{figure}[htbp]
    \begin{minipage}{0.333\textwidth}
        \includegraphics[width=\textwidth]{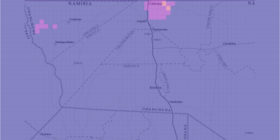}
    \end{minipage}\begin{minipage}{0.333\textwidth}
        \includegraphics[width=\textwidth]{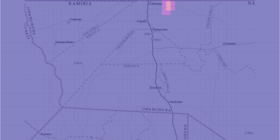}
    \end{minipage}\begin{minipage}{0.333\textwidth}
        \includegraphics[width=\textwidth]{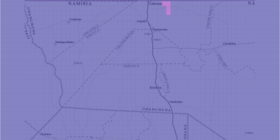}
    \end{minipage}

    \begin{minipage}{0.333\textwidth}
        \centering May 2011
    \end{minipage}\begin{minipage}{0.333\textwidth}
        \centering June 2011
    \end{minipage}\begin{minipage}{0.333\textwidth}
        \centering July 2011
    \end{minipage}

    \begin{minipage}{0.333\textwidth}
        \includegraphics[width=\textwidth]{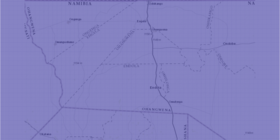}
    \end{minipage}\begin{minipage}{0.333\textwidth}
        \includegraphics[width=\textwidth]{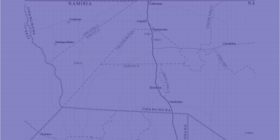}
    \end{minipage}\begin{minipage}{0.333\textwidth}
        \includegraphics[width=\textwidth]{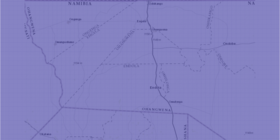}
    \end{minipage}

    \begin{minipage}{0.333\textwidth}
        \centering August 2011
    \end{minipage}\begin{minipage}{0.333\textwidth}
        \centering September 2011
    \end{minipage}\begin{minipage}{0.333\textwidth}
        \centering October 2011
    \end{minipage}

    \begin{minipage}{0.333\textwidth}
        \includegraphics[width=\textwidth]{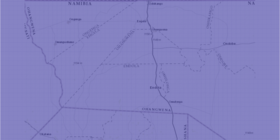}
    \end{minipage}\begin{minipage}{0.333\textwidth}
        \includegraphics[width=\textwidth]{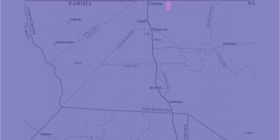}
    \end{minipage}\begin{minipage}{0.333\textwidth}
        \includegraphics[width=\textwidth]{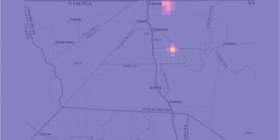}
    \end{minipage}

    \begin{minipage}{0.333\textwidth}
        \centering November 2011
    \end{minipage}\begin{minipage}{0.333\textwidth}
        \centering December 2011
    \end{minipage}\begin{minipage}{0.333\textwidth}
        \centering January 2012
    \end{minipage}

    \begin{minipage}{0.333\textwidth}
        \includegraphics[width=\textwidth]{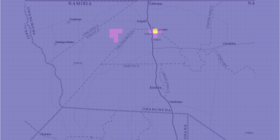}
    \end{minipage}\begin{minipage}{0.333\textwidth}
        \includegraphics[width=\textwidth]{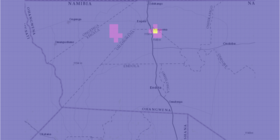}
    \end{minipage}\begin{minipage}{0.333\textwidth}
        \includegraphics[width=\textwidth]{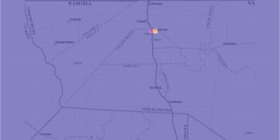}
    \end{minipage}

    \begin{minipage}{0.333\textwidth}
        \centering February 2012
    \end{minipage}\begin{minipage}{0.333\textwidth}
        \centering March 2012
    \end{minipage}\begin{minipage}{0.333\textwidth}
        \centering April 2012
    \end{minipage}

    \caption{\label{fig:months_13_24}Months 13--24. Plots showing $\P[\exp\{Y\}>1.5|\text{data}]$ i.e. the posterior probability that the relative risk is greater than 1.5. Colour key: \protect\includegraphics[width=1em]{col1.png} [0.0,0.2], \protect\includegraphics[width=1em]{col2.png} (0.2,0.4], \protect\includegraphics[width=1em]{col3.png} (0.4,0.6], \protect\includegraphics[width=1em]{col4.png} (0.6,0.8], \protect\includegraphics[width=1em]{col5.png} (0.8,1.0].}
\end{figure}

Figures \ref{fig:months_1_12} and \ref{fig:months_13_24} shows the probability that the relative risk exceeds 1.5; the former for months 1--12 and the latter for months 13--24. Month 25. Table \ref{tab:stparest} shows the fixed effects from the model, which show a significantly increase in risk of malaria with increasing EVI and LST; NDWI was not significant in this analysis. These results also show that there is little dependence between time points for these months.

\begin{table}[htbp]
    \centering
    \begin{tabular}{cccc}
          & 50\% & 2.5\% & 97.5\% \\ \hline
        Intercept & 0.165 & 0.116 & 0.253 \\
        EVI & 2.16 & 1.37 & 3.6 \\
        LST & 1.61 & 1.13 & 2.3 \\
        NDWI & 0.972 & 0.57 & 1.56 \\
        $\sigma$ & 2.62 & 2.11 & 3.2 \\
        $\phi$ (km) & 1.82 & 1.03 & 3.02 \\
        $a$ & 5.64$\times10^{-4}$ & 5.37$\times10^{-8}$ & 4.24$\times10^{-3}$ \\
    \end{tabular}
    \caption{\label{tab:stparest}Parameter effects from the spatiotemporal model: median and 95\% credible interval.}
\end{table}

\section{Application: Evaluating the Effect of Proposed Election Boundary Changes Around the Northern Powerhouse \label{sect:voting}}

In September 2016, the BBC reported that the House of Commons is considering reducing the number of UK Members from 650 to 600 \citep{BBC_parliament2016a,BBC_parliament2016b}. It is natural to seek to understand the potential impact these proposed (major) changes to the UK political constitution might have on voting patterns. In this section, we analyse data from the 2010 and 2015 general elections and look to evaluate the impact of the proposed boundary changes in the North West in an area around the City of Manchester. In this section, we use aggregated point processes to model and forecast the number of votes for the Conservative, Labour and Liberal Democrat parties.

\begin{figure}
    \begin{minipage}{0.5\textwidth}
        \begin{center}
            \includegraphics[width=\textwidth]{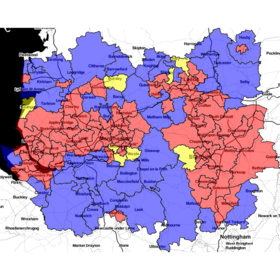}
        \end{center}
    \end{minipage}\begin{minipage}{0.5\textwidth}
        \begin{center}
            \includegraphics[width=\textwidth]{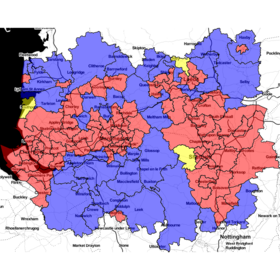}
        \end{center}
    \end{minipage}

    \begin{minipage}{0.5\textwidth}
        \begin{center}
            (a)
        \end{center}
    \end{minipage}\begin{minipage}{0.5\textwidth}
        \begin{center}
            (b)
        \end{center}
    \end{minipage}

    \caption{\label{fig:election_results} Showing voting results of 2010 (a) and 2015 (b) elections arond Manchester. Key: \protect\includegraphics[width=1em]{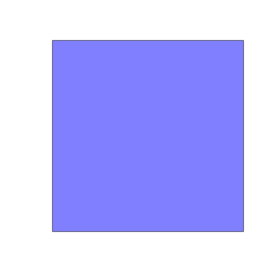} Conservative Party majority, \protect\includegraphics[width=1em]{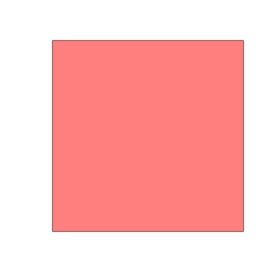} Labour  Party majority, \protect\includegraphics[width=1em]{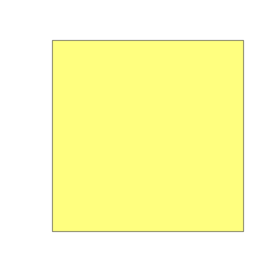}Liberal Democrat Party majority}
\end{figure}

The (arbirtary) area we chose to analyse extends from the west coast as far east as York; and with York also as the most northerly city, it extends as far south as Uttoxeter. This area includes the cities of Blackpool, Chester, Doncaster, Liverpool, Leeds, Manchester, Mansfield, Preston, Sheffield and York. We did not include the 2005 election results in our analysis because we were unable to obtain a reliable shapefile containing the results, though had we been able to, this would also have illustrated how our method works with changing boundaries over time. The results of the 2010 and 2015 general elections are shown in Figure \ref{fig:election_results}.

\begin{figure}
    \centering
    \begin{minipage}{0.6\textwidth}
        \begin{center}
            \includegraphics[width=\textwidth]{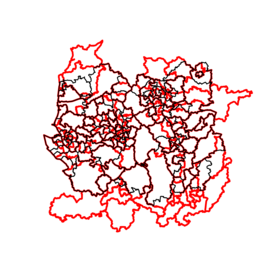}
        \end{center}
    \end{minipage}\begin{minipage}{0.4\textwidth}
        \begin{center}
            \includegraphics[width=\textwidth]{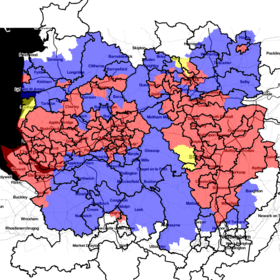}
        \end{center}
    \end{minipage}
    \caption{\label{fig:boundary_changes}Illustrating the changes to electoral ward boundaries. Left: black electoral ward boundaries as per the 2010 and 2015 elections. Red: proposed boundaries for the 2020 election. Right: showing results from the 2015}
\end{figure}

Figure \ref{fig:boundary_changes} shows the proposed changes to the boundaries in this area. Probably the most interesting feature in the left hand plot concerns the proposed changes to the three Liberal Democrat seats, including the seat of the former party leader, Nick Clegg. Under the present electoral boundary proposal, each of these three seats is proposed to be split into at least two further regions that at present have a majority Labour presence.

In terms of data available at the sub-aggregate (i.e. sub-electoral ward) level, we obtained 2010 population and deprivation data (the Index of Multiple Deprivation, IMD) at the Lower Super Output Level (LSOA) level from respectively the Office for National Statistics and the Department for Communities and Local Government. We included in our model the subset of IMD domains: income, health, crime and environment. Of the other three domains, barriers, education and employment, the latter two were highly correlated with income (0.88 and 0.92 respectively) and the barriers domain is difficult to interpert in the current context (it encompasses indicators for overcrowding, homelessness, access to owner-occupation (housing), road distance to GP surgery, road distance to store/supermarket, road distance to primary school and road distance to post office) \citep{IMD2010}. In addition to the deprivation measured, we included as indicators for the prevalence of certain societal attitudes in each electoral ward the proportion of people voting for the green party and for UKIP in the general election of 2015. Each of the IMD and societal covariates was rasterised to the computational grid and having done this, we standardised each of the covariates to allow us to compare which factors were the most important by examining the regression coefficients.

We used the same grid-level spatiotemporal model as per the malaria example above, and again used Gaussian priors for the log of $\sigma$, $\phi$ and $\theta$:
\begin{eqnarray*}
   \log\sigma&\sim&\N(0,0.3^2)\\
   \log\phi&\sim&\N(\log(5000),0.3^2)\\
   \log\theta&\sim&\N(-1,1)
\end{eqnarray*}
and independent $\N(0,10000^2)$ priors for each component of $\beta$. We used a non-extended $128\times128$ grid and ran each sampler for 1,000,000 iterations, discarding the first 50,000 as burnin and thereafter retaining every 950th for analysis.
% betaprior <- betapriorGauss(mean=0,sd=10000)
% omegaprior <- omegapriorGauss(mean=-1,sd=1)
% etaprior <- etapriorGauss(mean=log(c(1,5000)),sd=c(0.3,0.3))

In this example, we analysed the number of votes divided by 100 for each of the Conservative, Labour and Liberal democrat parties using three separate aggregated point process models. We divided the number of votes by 100 because mixing was slow for the full number of votes (in 2015, there were in total 1,611,910 Conservative, 2,380,969 Labour and 339,985 Liberal Democrat votes).

Table \ref{tab:elect_results} shows the paramater estimates from our model for each of the parties; the difference between an estimated effect size for covariate $k$, $\beta_k$ and 1 i.e. $|1-\beta_k|$ is an indicator of the size of the effect the $\beta$s can be compared both within and between models. The estimated effects, $\beta_k$ are essentially relative risks (RR): values above 1 indicate that greater deprivation in a particular domain is associated with more votes and values below 1 indicate that lesser deprivation in a particular domain is associated with fewer votes.

With these interpretations in mind, and in reverse order of effect size: a greater number of votes cast for the Conservative party was related to lower income deprivation RR 0.763 (0.657,0.899), lower crime deprivation RR 0.791 (0.725,0.871), lower health deprivation RR 0.838 (0.736,0.918) but not related to environment deprivation 0.99 (0.904,1.07); a greater number of votes cast for the Labour party was associated with greater health deprivation RR 1.2 (1.03,1.32) and greater crime deprivation RR 1.18 (1.07,1.32) and there was inconclusive evidence for there being greater numbers of votes in areas that were less income deprived RR 0.955 (0.869,1.03) and in areas with greater environment deprivation RR 1.04 (0.987,1.09);  a greater number of votes cast for the Liberal Democrat Party was associated with less income deprivation RR 0.545 (0.327,0.792), but greater crime deprivation RR 1.44 (1.01,1.81) and there was inconclusive evidence for there being greater numbers of votes in areas of lower environment deprivation RR 0.835 (0.676,1.01) and higher health deprivation RR 1.13 (0.838,1.63).

Clearly, we expect there to be proportionally fewer votes for each of the three main parties in the presence of greater numbers of voters for other parties and this is reflected in the coefficients for proportion Green and proportion UKIP being below 1 across all parties. Of the remaining coefficients the value of $\sigma$ is higher for the Liberal Democrats, which indicates voter support is spatially more sporadic.

\begin{table}[htbp]
    \centering
    \tiny
    \begin{tabular}{l|l|l|l}
          & coefCON & coefLAB & coefLIB \\ \hline
        Intercept & 3.09$\times10^{-3}$ (2.86$\times10^{-3}$,3.37$\times10^{-3}$) & 1.98$\times10^{-3}$ (1.82$\times10^{-3}$,2.19$\times10^{-3}$) & 8.92$\times10^{-4}$ (6.56$\times10^{-4}$,1.2$\times10^{-3}$) \\
        Income & 0.763 (0.657,0.899) & 0.955 (0.869,1.03) & 0.545 (0.327,0.792) \\
        Health & 0.838 (0.736,0.918) & 1.2 (1.03,1.32) & 1.13 (0.838,1.63) \\
        Crime & 0.791 (0.725,0.871) & 1.18 (1.07,1.32) & 1.44 (1.01,1.81) \\
        Environment & 0.99 (0.904,1.07) & 1.04 (0.987,1.09) & 0.835 (0.676,1.01) \\
        Prop. Green & 0.933 (0.902,0.968) & 0.936 (0.91,0.969) & 0.976 (0.904,1.04) \\
        Prop. UKIP & 0.934 (0.88,0.996) & 0.941 (0.88,0.991) & 0.886 (0.745,1.04) \\
        $\sigma$ & 0.446 (0.391,0.518) & 0.427 (0.391,0.485) & 1.2 (1.09,1.33) \\
        $\phi$ & 4279 (3140,5778) & 5287 (3966,7323) & 7391 (6059,9274) \\
        $a$ & 4.99$\times10^{-3}$ (1.9$\times10^{-2}$,1.89$\times10^{-5}$) & 3.65$\times10^{-3}$ (2.07$\times10^{-2}$,2.82$\times10^{-10}$) & 3.17$\times10^{-3}$ (1.62$\times10^{-2}$,1.48$\times10^{-6}$) \\
    \end{tabular}
    \caption{\label{tab:elect_results}Parameter effects from the election model.}
\end{table}

\begin{figure}
    \centering
    \includegraphics[width=0.7\textwidth]{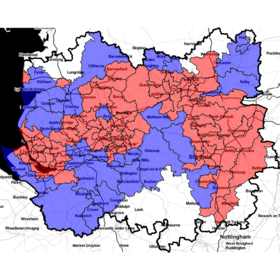}
    \caption{\label{fig:predicted}Predicted party majorities under new boundaries, see text for details on how this was calculated. Key: \protect\includegraphics[width=1em]{colCON.png} Conservative Party majority, \protect\includegraphics[width=1em]{colLAB.png} Labour Party majority.}
\end{figure}

Figure \ref{fig:predicted} and Table \ref{tab:forecast} show the forecast majority, based on the largest number of predicted votes. These predictions were computed in a two stage process. Firstly, for each retained MCMC sample, we can sample from the model given those parameters which yields a predicted number of votes for each party in each computational grid cell. Then we aggregate the predicted cell counts in each new electoral ward by summing the cell-level counts, weighted by the proportion of each cell contained in the electoral ward. For each party and region we therefore obtain 1000 realisations of the predictive distribution for the number of votes and declare the victor to be the party with the largest predicted median number of votes. Table \ref{tab:forecast} gives the number and proportion of seats for each party in 2010, 2015 and predicted for 2020 under the boundary changes. The results show that Labour, at least in this area of the country, is set to benefit from the proposed changes at the expense of the Liberal Democrats.

\begin{figure}
    \begin{minipage}{0.333\textwidth}
        \includegraphics[width=\textwidth]{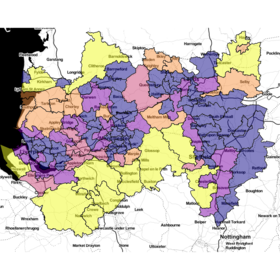}
    \end{minipage}\begin{minipage}{0.333\textwidth}
        \includegraphics[width=\textwidth]{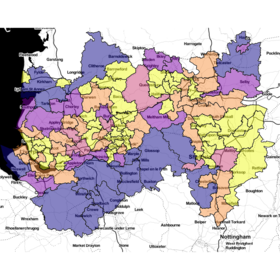}
    \end{minipage}\begin{minipage}{0.333\textwidth}
        \includegraphics[width=\textwidth]{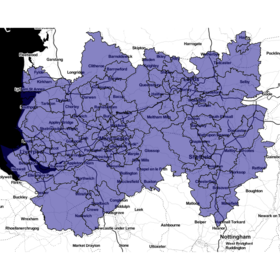}
    \end{minipage}
    \caption{\label{fig:winprob} Left to Right, illustrating the probability of a Conservative, Labour or Liberal Democrat win. Colour key: \protect\includegraphics[width=1em]{col1.png} [0.0,0.2], \protect\includegraphics[width=1em]{col2.png} (0.2,0.4], \protect\includegraphics[width=1em]{col3.png} (0.4,0.6], \protect\includegraphics[width=1em]{col4.png} (0.6,0.8], \protect\includegraphics[width=1em]{col5.png} (0.8,1.0].}
\end{figure}

\begin{table}[htbp]
    \centering
    \begin{tabular}{cccc}
          & 2010 & 2015 & 2020 \\ \hline
        CON & 32(0.27) & 33(0.28) & 28(0.27) \\
        LAB & 78(0.66) & 82(0.69) & 74(0.73) \\
        LIB & 8(0.07) & 3(0.03) &   \\
    \end{tabular}
    \caption{\label{tab:forecast} Predicted number of seats in the highlighted areas in Figure \ref{fig:predicted}.}
\end{table}
One thing not clear from Figure \ref{fig:predicted} is the uncertainty surrounding these predictions. Therefore, we also computed the probability of a victory for each of the parties as the empirical proportion of times each party took the majority share under each of the 1000 simulated aggregated vote counts; these are shown in Figure \ref{fig:winprob}.

The purpose of this analysis was to illustrate how our methods can be used to deliver inference on a dataset in which predicting the counts on a new partition is important. A more rigorous analysis of the data might seek to model competition for votes between the parties and in particular, some care should be taken in interpreting the predicted probabilities in Figure \ref{fig:winprob}, as the main assumption is independence in the predicted number of votes. One way of extending this model would be to use a multivariate spatiotemporal LGCP to jointly model the counts for all parties simultaneously.

\section{Discussion\label{sect:discussion}}

In this article, we have introduced a method for delivering continuous inference from aggregated data in which the aggregation units potentially overlap, are uncertain and change over time. We have investigated how inferences are affected by differing amounts of overlap and have illustrated our methods on two real-world examples. Our proposed method could find wide application in the modelling of aggregated count data, since these types of dataset are very common. It is also commonplace for administrative boundaries to change over time, and though we have not analysed such a dataset in the present article, by utilising a common inferential grid at each time point our method offers an elegant solution to this issue. The main strength of our proposed approach is that we have attacked the problem of inference for aggregated data from the modelling perspective: we advocate modelling the stochastic process that is hypothesised to have generated the data, rather than by modelling the data as it happens to have been stored.

We have used the term `continuous' to describe the type of inference we get from our proposed model and method of analysis, even though in truth we deliver inference on merely a different discretisation of space. We believe the use of the word `continuous' can be justified on the basis that we are at liberty, computationally permitting, to make our inferential grid as fine as we like, and in the limit, we approach continuity.  The same cannot of course be said of the covariates. In most cases, covariates can only be measured on a finite division of space (e.g. a raster image, or a multipolygon), thus our ability to make truly continuous inferences is in some sense dictated by the level of discretisation of the covariates. An interesting area for future research would be to introduce additional model hierarchies to handle mapping of the covariates in a smooth manner onto the computational grid. A second reason for describing our methods as `continuous' is that we have employed non-Markovian models for spatial correlation in the field $Y$; the extension of our methods to Markov models or low-rank models of spatial/spatiotemporal correlation is trivial.

One thing we have not fully explored in the current article is the extent to which uncertainty in stochastic boundary definition affects our ability to deliver inference for key model parameters. As mentioned in Section \ref{sect:uncertain}, it is clear that in this situation, our ability to estimate other model parameters will be compromised. We conjecture that there is no general solution to this issue, and much of our ability to estimate parameters will depend on the available data and suitability of the model.

Another simple extension of our proposed modelling framework is where we have a mix of point-referenced and aggregated data. Recall that in the Gibbs scheme, we alternately draw from: $\pi(\beta,\eta,Y|N,T_{1:m})$  and $\pi(N|\beta,\eta,Y,T_{1:m})$, denoting by $X$ the cell counts implied by the point-referenced data, we instead draw from $\pi(\beta,\eta,Y|N,X,T_{1:m})$  and $\pi(N|\beta,\eta,Y,X,T_{1:m})$. The first of these densities simplifies to $\pi(\beta,\eta,Y|N,X)$ and sampling proceeds as before (the cell counts just being $N+X$) and the second density also retains its form as a multinomial as long as there is no spatial bias in the sampling of $X$ \citep{diggle2010a}.

The main disadvantage of our method is currently computational cost. However, many aspects of our proposed algorithm can be performed in parallel, including the fast Fourier transform and obviously matrix multiplications. We have already developed \proglang{CUDA} GPU code to implement inference for the point-referenced spatial log-Gaussian Cox process with some success: the parallel implemetation being just over 10--30 times faster than our \proglang{R} implementation (depending on the size of the inferential grid and GPU availability). We are currently developing \proglang{CUDA} code and an \proglang{R} interface to deliver inference for the models described in the present article.

\section*{Acknowledgements}

The authors would like to thank Davis Mumbengegwi and Joyce Auala from The University of Namibia for use of the malaria case data in Section \ref{sect:namibia} and Emanuele Giorgi for providing feedback on a draft of the manuscript.

In our \code{R} implementation, the packages \pkg{sp} and \pkg{rgeos} have been particularly important for handling spatial geometries and computing intersections between polygons \citep{pebesma2005,bivand2013,bivand2016}.

The IMD and LSOA data in this article are subject to the Open Government License: \url{http://www.nationalarchives.gov.uk/doc/open-government-licence/}.

Contains National Statistics data \copyright Crown copyright and database right 2001, 2004, 2007

Contains Ordnance Survey data \copyright Crown copyright and database right 2001, 2004, 2007

Map tiles by Stamen Design, under CC BY 3.0. Data by OpenStreetMap, under ODbL

\bibliographystyle{chicago}
\bibliography{bibliography}

\end{document}